# A simple model for predicting crystallization and melting temperatures, and its implications for phase transitions in confined volumes.

**Cooper, Sharon; Nicholson, Catherine; Liu, Jian**

Chemistry Department, Science Laboratories, Durham University, South Road, Durham, DH1 3LE, UK.

## ABSTRACT

We present a simple unifying model for crystallization and melting temperatures by showing that homogeneous nucleation and phase transformations driven by thickening of pre-existing surface layers are limiting conditions of the more general heterogeneous nucleation case. Furthermore, to a first approximation all these processes can be described by an extended classical nucleation theory. The model can also be applied to phase transition temperatures in confined volumes, provided reliable values for the interfacial tensions within the systems are determinable. The expected melting and crystallization temperature for any transformation pathway can then be predicted.

## I. INTRODUCTION

Crystallization and melting temperatures of materials in confined volumes can vary extensively from those observed in the bulk phases. Hence it is important that theories are derived to model these cases, particularly as the findings are pertinent to a diverse range of areas including nanomaterial production, rock weathering and oil recovery. The rate of crystallization and melting at a particular temperature is determined primarily by the size of the energy barrier to the creation of the new phase. For most substances, normal alkanes being a notable exception, surface melting occurs at a lower temperature than the bulk, and hence bulk melting occurs by thickening of this layer without the need for superheating above the equilibrium melting temperature, $T_0$. For small particles, this results in melting below the equilibrium melting temperature, with the melting point depression often modelled [1,2] by the Gibbs-Thomson equation,

$$\Delta T = T_0 - T_m = (2\gamma v T_0)/R\Delta_{fus}H \qquad (1)$$

where $T_m$ is the melting temperature, $\gamma$ is the melt-crystal interfacial tension, $v$ is the liquid molecular volume, $R$ is the particle radius and $\Delta_{fus}H$ is the enthalpy of fusion. Theoretical treatments have shown [3-5], however, that the Gibbs-Thomson equation actually represents to the first approximation [6] the upper bound for melting of small solid particles when a surface liquid layer is present. This is because the Gibbs-Thomson equation gives the condition for the unstable equilibrium between the solid particle and the melt, and so represents the melting transformation pathway for which the energy barrier is zero. This is qualitatively easy to understand from the following. Melting of a small particle with a surface liquid layer is accompanied by an energy decrease due to the reduction in the interfacial area between the solid and the melt, and an energy increase due to the production of the thermodynamically less stable phase if the temperature is below the equilibrium melting temperature, $T_0$. For the solid particle at equilibrium with the melt, for which the Gibbs-Thomson equation holds, these two energies cancel. As this solid particle melts, however, its surface area to volume ratio increases, and so the energy term due to the interfacial area reduction dominates, resulting in an increasing energy lowering, and hence no energy barrier to melting. The thermal energy available to surmount energy barriers would then be expected to allow melting of the particle below this temperature.

A lower bound of the melting point has also been derived [7], based on the criterion that the melting of the particle must be thermodynamically feasible, i.e. the change in the free energy must be ≤ 0. This criterion is satisfied when $\Delta T = (3\gamma v T_0)/R\Delta_{fus}H$. Hence

$$T_0\left[1-\frac{3\gamma v}{R\Delta_{fus}H}\right] \leq T_m \leq T_0\left[1-\frac{2\gamma v}{R\Delta_{fus}H}\right]. \qquad (2)$$

For crystallization, the absence of a pre-existing solid layer means that the phase transformation must typically occur via formation of a stable crystalline nucleus, which then grows to become the new bulk phase. The energy barrier to the creation of the stable nucleus results in supercooling being required before the new phase appears. The size of the crystallization energy barrier can be modelled using classical nucleation theory. Recently, we have adapted [8] classical nucleation theory to account for crystallization within confined volumes. We now extend this theory and show that melting via thickening of a pre-existing surface layer is a limiting case of this more general extended nucleation theory. Thus, a single melting temperature can be assigned for a particular particle size, instead of a temperature range between the upper and lower temperature bounds. This allows crystallization and melting temperatures in any system to be described by our simple model.

The aim of this paper is to outline our model and to detail its uses and limitations. In particular, our mesoscopic model uses interfacial tensions to describe the interactions between surfaces, and as such does not describe the molecular interactions between surfaces needed to explain the occurrence, or otherwise, of premelting. Many detailed atomistic approaches have already been described to account for this phenomenon in confined geometries [9-12]. The interfacial tension values used in our model are necessarily isotropic, since the crystallite orientation on the surface is not known [13], and as the confining volume decreases in size, the interfacial tension values should be expected to deviate from bulk values. Despite these limitations, our model is useful because it provides analytical expressions for melting and crystallization temperatures in confined volumes that represent an improvement over those based on the Gibbs-Thomson equation and unmodified classical nucleation theory. These analytical expressions can readily be applied to experimental data and have recently enabled the first direct measurement of the critical nucleus size for ice crystallization in microemulsions [14].

## II. THEORETICAL MODEL

### A. Classical nucleation theory

The formation of any new phase from a bulk parent phase requires the creation of an interface between the two phases, which requires work. Hence there exists an energy barrier, $\Delta G^*$, to the formation of the new phase. In classical nucleation theory [15], the value of $\Delta G^*_{hom}$ for homogeneous nucleation, i.e. nucleation within the bulk parent phase, is calculated as follows. The change in the Gibbs free energy, $\Delta G_{hom}$, in forming a spherical nucleus of $i$ molecules from isolated molecules is given by the sum of favourable volume and unfavourable surface area terms, i.e.:

$$\Delta G_{hom} = -i\Delta\mu + \Sigma\gamma A = -\frac{4\pi r^3 \Delta\mu}{3v} + 4\pi r^2 \gamma \qquad (3)$$

where $\Delta\mu$ = the supersaturation, $\gamma$ = interfacial tension between the nucleus and the surrounding medium, $r$ = radius of the nucleus and $v$ = molecular volume of the isolated molecule. The supersaturation is the driving force for condensation and is given by:

$$\Delta\mu = \Delta_{fus}H\Delta T/T_0. \tag{4}$$

The barrier to nucleation, $\Delta G^*_{hom}$, is found from the maximum in $\Delta G_{hom}$ by setting $dG_{hom}/dr = 0$. This condition is satisfied when $r = r^* = 2\gamma v/\Delta\mu$, i.e. by the Gibbs-Thomson equation, and we find that:

$$\Delta G^*_{hom} = \frac{16}{3\Delta\mu^2}\pi\gamma^3 v^2. \tag{5}$$

B. Extension of classical nucleation theory to curved substrates

This approach can be extended to provide the energy barrier for heterogeneous nucleation, on or within a spherical substrate, of radius $R$, as shown in Figure 1. In this case, the Gibbs free energy change, $\Delta G_{het}$, is given to a first approximation [6] by [8,16]

$$\Delta G_{het} = -\frac{4}{3v_c}\pi r^3\Delta\mu[(f(\theta+\phi)-(R/r)^3 f(\phi)] + 2[1-\cos(\theta+\phi)]\pi r^2\gamma + 2(1-\cos\phi)\pi R^2(\gamma_2-\gamma_1)$$
$$\tag{6}$$

where $\theta$ = contact angle between the nucleus and the spherical substrate, $\phi$ is the angle between the spherical substrate and the plane located at the perimeter of the cluster, and $\gamma_1$ and $\gamma_2$ are the interfacial tensions between the substrate and bulk phase, and substrate and nucleus, respectively. Please note that for crystallization within or upon confined volumes, the Helmholtz free energy is more appropriate than the Gibbs free energy, owing to the Laplace pressure difference across curved surfaces. However we retain use of the Gibbs free energy throughout to emphasise the connection with the classical nucleation theory, because we are assuming to a first approximation that the phases are incompressible and that there is no volume change on phase transformation.

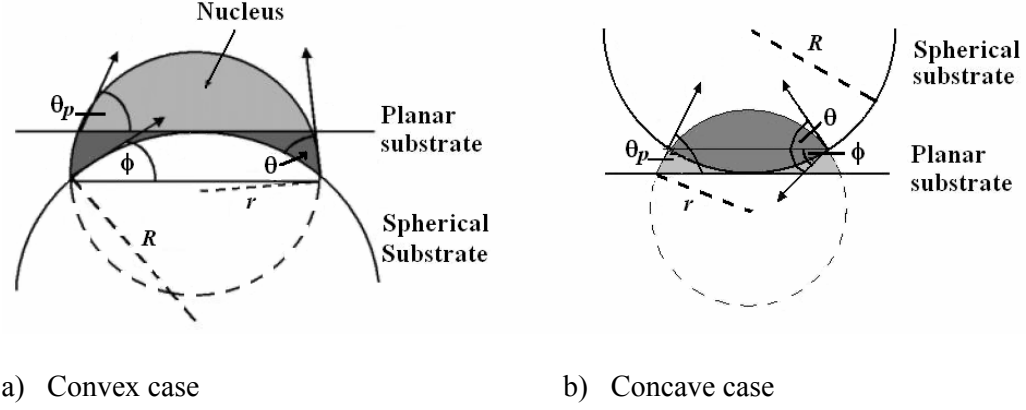

a) Convex case    b) Concave case

*FIG 1. Schematic diagram describing nucleation a) upon and b) within a droplet of radius, R. In a) the combined light and dark grey regions depict the nucleus on the convex curved substrate, whereas in b) the dark grey regions depict the nucleus on the concave surface.*

From Young's equation: $\cos\theta = (\gamma_1 - \gamma_2)/\gamma$, so

$$\Delta G_{het} = -\frac{4}{3v_c}\pi r^3 \Delta\mu[(f(\theta+\phi) - (R/r)^3 f(\phi)] + 2[1 - \cos(\theta+\phi)]\pi r^2 \gamma - 2\cos\theta(1-\cos\phi)\pi R^2 \gamma$$

(7)

The maximum in $\Delta G_{het}$ gives the barrier to nucleation, $\Delta G^*_{het}$, and again this condition is satisfied [16] when $r = r^* = 2\gamma v/\Delta\mu$.

Hence:

$$\Delta G^*_{het} = \frac{32}{3\Delta\mu^2}\pi\gamma^3 v^2[-f(\theta+\phi) + (R/r^*)^3 f(\phi)] + \frac{8}{\Delta\mu^2}\pi\gamma^3 v^2[\{1 - \cos(\theta+\phi)\} - \cos\theta(1-\cos\phi)(R/r^*)^2]$$

Rearrangement and simplification gives:

$$\Delta G^*_{het} = \frac{\Delta G^*_{hom}}{2}\left[1 - \cos^3(\theta + \phi) + 4x^3 f(\phi) - 3x^2 \cos\theta(1 - \cos\phi)\right] \quad (8)$$

where $x = R/r^*$. This equation applies to both nucleation upon a spherical substrate, to give the convex case shown in Fig 1a, or within one to give the concave case shown in Fig 1b. For the concave system, though, $R$ and $\phi$ have negative values.

Greater insight into equation (8) is provided by introducing the variable, $y$, where:

$$y = \pm(x^2 - 2x\cos\theta + 1)^{0.5} \quad (9)$$

The positive root is used for a nucleus on a convex surface, whereas the negative root applies for the concave case. The variable $|y|$ represents the third side of a triangle whose other two lengths are 1 and $|x|$, with the angle between the sides of 1 and $|x|$ being $\theta$ for the convex case and (180°-$\theta$) for the concave case. The angle between the sides of length $|y|$ and $|x|$ is then $|\phi|$, see Figure 2.

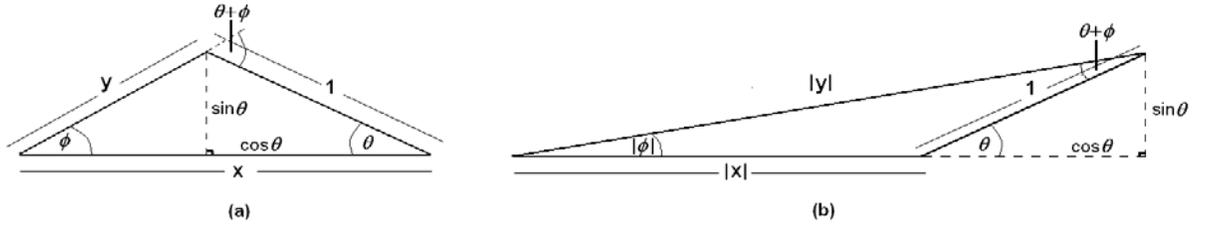

FIG 2. Diagram showing the geometric relationships between x, y, $\theta$, $\phi$ and ($\theta+\phi$) for

a) the convex system and b) the concave system.

Using this we find that $\cos\phi = (x - \cos\theta)/y$ and $\cos(\theta + \phi) = (x\cos\theta - 1)/y$. After substitution and simplification, equation (8) then becomes:

$$\Delta G^*_{het} = \frac{\Delta G^*_{hom}}{2}\{1 - 3x^2\cos\theta + 2x^3 + y(1 + x\cos\theta - 2x^2)\} = \Delta G^*_{hom} f(\theta_p) \quad (10)$$

where $\theta_p$ is the angle between the critical nucleus and the corresponding planar surface tangential to the curved surface, see Figure 1. Hence the heterogeneous nucleation barrier for any size concave or convex substrate can be predicted by determining the value of $\theta_p$. $\theta_p$ is given simply by

$$\cos\theta_p = x - y. \tag{11}$$

A proof of the equivalence of the RHS of equations (8) and (10) is given in the Appendix. Equation (10) shows that at a given temperature, and hence constant $\Delta G^*_{hom}$ value, the value of $\Delta G^*_{het}$ depends only upon the $\theta_p$ value. Consequently, in Figure 3, all the spherical substrates depicted result in the same $\Delta G^*_{het}$ value. This can be rationalised as follows. For nucleation on a concave surface, the critical nucleus volume, $V^*$, is reduced compared to the planar case, and hence fewer molecules need to cluster together to form the critical nucleus. However this effect is negated by the greater contact angle, $\theta$, compared to $\theta_p$, which means that more work is required to create unit area of the nucleus-substrate interface and so the mean energy increase on addition of a molecule to the sub-critical nucleus is larger, see Figure 4. In contrast, for nucleation on a convex substrate, $V^*$ is increased compared to the planar case, but $\theta$ is decreased.

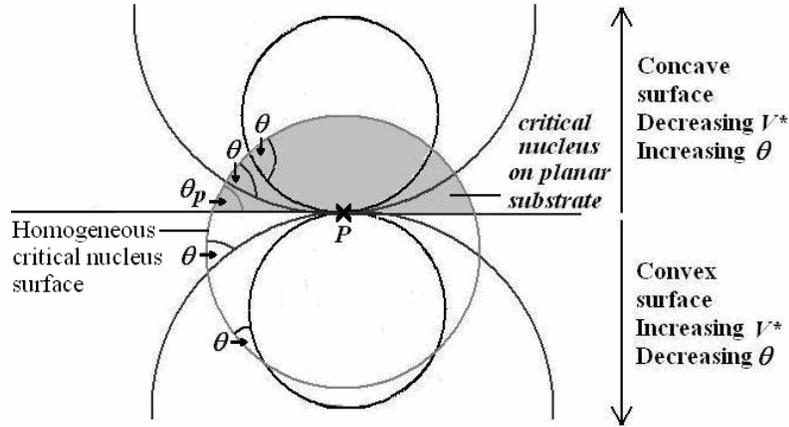

*FIG 3. Schematic diagram showing that for a given supersaturation, and hence critical nucleus radius, r\*, all surfaces through point P that cross the homogeneous critical nucleus surface [28] produce the same $\Delta G^*_{het}$ value for nucleation, since they all have the same $\theta_p$ value.*

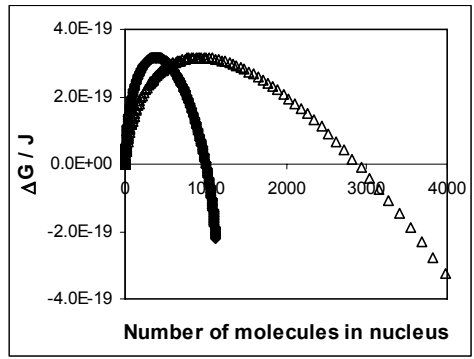

*FIG 4. Graph of $\Delta G_{het}$ vs. number of molecules in an ice nucleus for a concave (filled squares) and convex (unfilled triangles) droplet of the same radius, 2.1 nm, with the same r\* value of 2.2 nm and the same $\theta_p$ value of 88.12° when r = r\*. This corresponds to systems with a contact angle of 120° for the concave case, and 55.6° for the convex one. A value of $\gamma = 33$ mN m$^{-1}$ [18] has also been used in equation (7).*

Having obtained the energy barrier to nucleation, $\Delta G^*_{het}$, the highest temperature, $T_c$, at which crystallization should be observable can be found. This is achieved by setting the nucleation rate, $J$, at $T_c$ to be 1 cm$^2$ s$^{-1}$, where $J = A\exp(-\Delta G^*_{het}/kT_c)$ and $A$ is the pre-exponential factor, which is assumed to be constant [17]. Hence: $(\Delta G^*_{het}/kT_c) = \ln A$, i.e.

$$\Delta T^3 - \Delta T^2 T_0 + \frac{16\pi\gamma^3 v^2 T_0^2 f(\theta_p)}{3k\Delta_{fus}H^2 \ln A} = 0. \quad (12)$$

Solving equation (12) for $T_c$, where $T_c \le T_0$ we obtain:

$$T_c = \frac{T_0}{3}\{2 + \cos X - (3^{0.5}\sin X)\} \quad (13)$$

where $X = \frac{1}{3}\arccos\left[1 - \frac{72\pi\gamma^3 v^2 f(\theta_p)}{k\Delta_{fus}H^2 T_0 \ln A}\right]$.

The second solution to equation (12) provides the expected melting temperature, $T_m$, for a nucleation-based melting transition, which would be required in the absence of a surface liquid layer. $T_m$ is given by:

$$T_m = \frac{T_0}{3}\{2 + \cos X + (3^{0.5}\sin X)\}. \quad (14)$$

The third solution, $T_c = (2T_0/3)\{1 - \cos X\}$ represents the non-physical case where $T_c$ is approaching 0 K and so the critical nucleus contains only one molecule and the energy barrier is vanishingly small. Crystallization would not be observed at this temperature, as freezing at the higher $T_c$ (or vitrification in the case of a sufficiently rapid quench) would have already occurred prior to this. Equations (13) and (14) are valid for nucleation on any shaped substrate, provided the appropriate geometric factor, $g$, is included in determining $X$, i.e. $X = \frac{1}{3}\arccos\left[1 - \frac{72\pi\gamma^3 v^2 g}{k\Delta_{fus}H^2 T_0 \ln A}\right]$.

For a spherical substrate, $g = f(\theta_p)$, whereas for a planar substrate $g = f(\theta)$. So for any concave shaped substrate we would expect $f(\theta_p) \leq g \leq f(\theta)$, whilst for any convex-shaped substrate we would have $f(\theta) \leq g \leq f(\theta_p)$.

Figures 5a and b show the predicted variation in $f(\theta_p)$ and the ice crystallization temperatures, respectively, as a function of substrate radius, $R$, for different contact angle systems. The ice crystallization temperatures have been determined using the standard values [18,19] of $\gamma = 33$ mN m$^{-1}$, $\Delta_{fus}H = 6010$ J mol$^{-1}$ and $A = 3 \times 10^{40}$ cm$^2$ s$^{-1}$. For the convex case, $y$ and $x$ are positive and $y \geq x$-cos $\theta$, so equation (11) shows that $\theta_p \geq \theta$. For the concave case, $y$ and $x$ are negative and $|y| \geq |x| + \cos\theta$, so $\theta_p \leq \theta$. Consequently, nucleation is easier on a concave surface compared to a convex one with the same $\theta$ value, and the difference becomes greater the smaller the magnitude of $x$. Indeed, for the concave case, as $|x|$ approaches zero, $y$ tends to -1, and $\cos\theta_p$ tends to 1, i.e. $\theta_p$ tends to 0°. Thus we would expect the barrier to nucleation to disappear for small enough droplet sizes, irrespective of the nucleus-droplet contact angle. In contrast, as $|x|$ tends to zero for the convex case, $y$ tends to 1 and $\cos\theta_p$ tends to -1, so $\theta_p$ tends to 180°. Consequently, $T_c$ is predicted to tend to the value expected for homogeneous nucleation for crystallization upon sufficiently small substrates, irrespective of the nucleus-substrate contact angle.

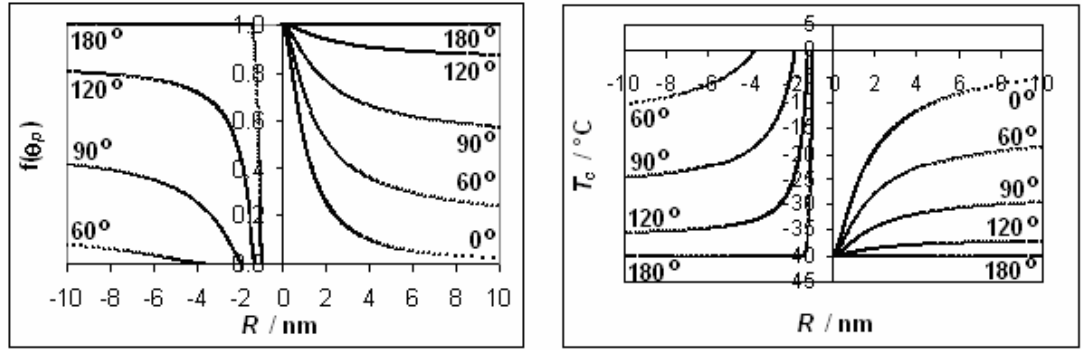

(a)                                               (b)

*FIG 5. The predicted variation in a) $f(\theta_p)$ and b) the ice crystallization temperatures, respectively, as a function of R for different contact angle systems. Negative and positive R values relate to crystallization within, and upon, spherical droplets, respectively.*

### C. Extension of classical nucleation theory to phase transitions induced by a reduction in interfacial energy

For the convex case, which describes crystallization upon droplets, $dT_c/dR$ tends to zero as $R$ approaches zero, and hence $T_c$ just tends to the homogeneous $T_c$ value. However for the concave case, it is clear from Figure 5b that $dT_c/dR$ does not tend to zero for small droplets when the nucleation barrier becomes vanishingly small. This suggests that for droplet sizes smaller than this, $T_c > T_0$. In fact, this situation is one of the general class of phase transformations driven not by the supersaturation of the parent phase, but *by the reduction in interfacial energy that arises*. In this situation, the reduction in interfacial energy upon nucleus growth is sufficiently large that it drives the phase transformation even though it produces the thermodynamically disfavoured phase, and so $\Delta\mu$ and $r^*$ are negative. Hence this case can be modelled for crystallization within droplets by determining $\theta_p$ values using positive $x$ values, as $R$ and $r^*$ are both negative now. The melting of small droplets with a surface liquid layer provides a well known example of this class of behaviour, for which it is known that $T_m < T_0$. In particular, the assignment of lower and upper bounds to the melting temperature reveals that $T_0[1-\{(3\gamma v T_0)/(R\Delta_{fus}H)\}] \leq T_m \leq T_0[1-\{(2\gamma v T_0)/(R\Delta_{fus}H)\}]$.
Recent papers have formulated [3-5] the energy change on thickening of the surface layer by which melting can proceed. To a first approximation [6], this energy change for melting of spherical droplets with a surface liquid layer is given by:

$$\Delta G_{surface\_layer} = -\frac{4}{3v}\pi\Delta\mu[(R^3-r^3)] + 4\pi r^2\gamma - 4\pi R^2\gamma \tag{15}$$

or

$$\Delta G_{surface\_layer} = -\frac{4}{3v}\pi\Delta\mu r^3[(x^3-1)] + 4\pi r^2\gamma(1-x)$$

where $R$ is the droplet radius, $x = R/r$ and $R-r$ is the thickness of the surface layer. Melting/crystallization of the whole droplet occurs when $r$ goes to zero, and the energy barrier for this process is given by $d\Delta G_{surface\_layer}/dr = 0$, with constant $R$ for melting of a given droplet, for which $r^* = -2\gamma v/\Delta\mu$. Substituting this $r^*$ value into equation (15) gives

$$\Delta G^*_{surface\_layer} = \frac{16\pi\gamma^3 v^2}{3\Delta\mu^2}\left(1 - 3x^2 + 2x^3\right). \tag{16}$$

The predicted phase transformation temperature, $T_{trans}$, can then be found as for the nucleation case by setting $\left(\Delta G^*_{het}/kT_{trans}\right) = \ln A$. However, our equation (10) i.e. $\Delta G^*_{het} = \left(\Delta G^*_{hom}/2\right)\{1 - 3x^2\cos\theta + 2x^3 + y(1 + x\cos\theta - 2x^2)\} = \Delta G^*_{hom} f(\theta_p)$ reduces to equation (16) when $\theta = 0$, as then $\cos\theta = 1$ and $y = 1-x$. So equation (16) just represents a limiting case of our nucleation equation (10) [20].

Figure 6 shows the ice melting temperatures given by equation (12) using the values of $\gamma = 33$ mN m$^{-1}$, $\Delta_{fus}H = 6010$ J mol$^{-1}$ and $A = 3 \times 10^{40}$ cm$^2$ s$^{-1}$ for systems with surface liquid layers. It can be seen that the predicted melting temperatures fall between the upper and lower bound melting temperatures given by $T_0[1-\{(2\gamma v T_0)/(R\Delta_{fus}H)\}]$ and $T_0[1-\{(3\gamma v T_0)/(R\Delta_{fus}H)\}]$, for droplet sizes above a limiting value $R_{min}$, suggesting that for $R > R_{min}$, our extended nucleation theory can be used to predict accurate melting temperatures. Furthermore, homogeneous nucleation represents the limiting case of equation (10) when θ = 180º. Homogeneous crystallization will occur for systems that melt via thickening of a surface liquid layer and so the crystallization curve for this case has been included in Figure 6. Therefore, equation (12) is a unifying equation to describe melting and crystallization temperatures irrespective of whether the phase transformation has to occur by a volume driven nucleation mechanism, or whether it can occur in undersaturated systems by a lowering of the interfacial energy of the system. All that is required to predict the phase transformation temperature for the particular driving mechanism is the determination of $f(\theta_p)$ from the correctly signed values of $x$ and $y$ [21]. The appropriate signs of $x$, $y$, $R$ and $r^*$ to determine $\theta_p$ for a given phase transformation mechanism and substrate curvature are listed in Table 1, along with the appropriate root of equation (12) to determine the phase transformation temperature.

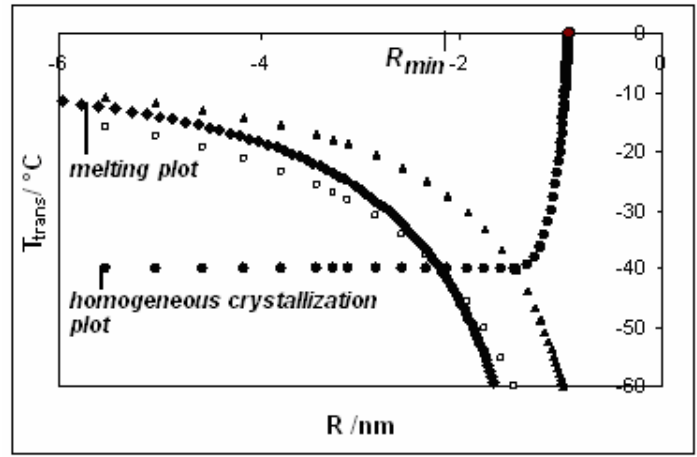

FIG 6. The predicted ice melting temperature (filled diamonds) as a function of droplet size for droplets with surface liquid layers. Note this melting curve falls between the upper (filled triangles) and lower (open circles) bounds for melting, until a droplet size $R_{min}$ is reached, where the melting and crystallization plots meet. The homogeneous ice crystallization curve (filled circles) is also shown.

**TABLE I** Parameter signs and equation numbers for predicting crystallization and melting temperatures in concave and convex systems.

| System | Process | Mechanism | $R$ | $y$ | $r$ | $x$ | Physical $\theta_p$? [21] | Equation No. for $T_{trans}$ |
|---|---|---|---|---|---|---|---|---|
| **Concave** | Cryst | Volume driven | -ve | -ve | +ve | -ve | √ | (13) |
| | Melt | Volume driven | -ve | -ve | +ve | -ve | √ | (14) |
| | Cryst | Interfacial area driven | -ve | -ve | -ve | +ve | X | (14) |
| | Melt | Interfacial area | -ve | -ve | -ve | +ve | X | (13) |

|        |      | driven           |     |     |     |     |   |      |
|--------|------|------------------|-----|-----|-----|-----|---|------|
|        |      |                  |     |     |     |     |   |      |
| **Convex** | Cryst | Volume driven | +ve | +ve | +ve | +ve | √ | (13) |
|        | Melt | Volume driven    | +ve | +ve | +ve | +ve | √ | (14) |

Equation (10) can of course be improved by removing the assumption of incompressible phases, and incorporating the effects of surface stress [22,23]. An exponential term to account for short-range interactions between the surfaces leading to ordering or disordering can also be included [3,4,24]. The use of bulk values for the interfacial tensions may also introduce errors as the confining volume decreases in size. For instance, the use of the bulk water-ice interfacial tension value was found to be inappropriate for sub 2 nm particles for ice crystallization in microemulsions [14]. Nevertheless, the model is attractive in its simplicity and because it is the first unifying approach to predicting melting and crystallization temperatures.

Figure 7 shows a schematic diagram outlining the different mechanisms for phase transformations within confined volumes. In particular, Figure 7a depicts the critical nucleus for heterogeneous nucleation within a droplet with $\theta > 90°$; the dashed outline represents growth of the nucleus, and this is favoured for supersaturated systems since the increased volume of the thermodynamically stable phase outweighs the unfavourable increase in the interfacial energy term arising from the increased interfacial area. In contrast, Figure 7b depicts the case where $r^*$ is negative, which can occur for sufficiently small $R$ when $\theta < 90°$. The phase transformation is now driven by the reduction in the interfacial energy of the system. The nucleus can grow in undersaturated systems because growth, shown by the dashed line, results in a decrease in the interfacial area and hence the interfacial energy of the system, and this outweighs the unfavourable increase in volume of the thermodynamically disfavoured phase. Figures 7c and 7d show the limiting cases of phase transformations occurring by homogeneous nucleation for a supersaturated system and by thickening of a pre-existing wetting layer for an undersaturated system, respectively. Again it can be seen that the nucleation mechanism requires a supersaturated system in order to compensate for the increased interfacial energy that arises from the nucleus growth shown by the dashed line, whereas if the phase transformation arises from thickening of a pre-existing wetting layer, then this can occur in undersaturated systems since the interfacial energy of the system decreases with nucleus growth.

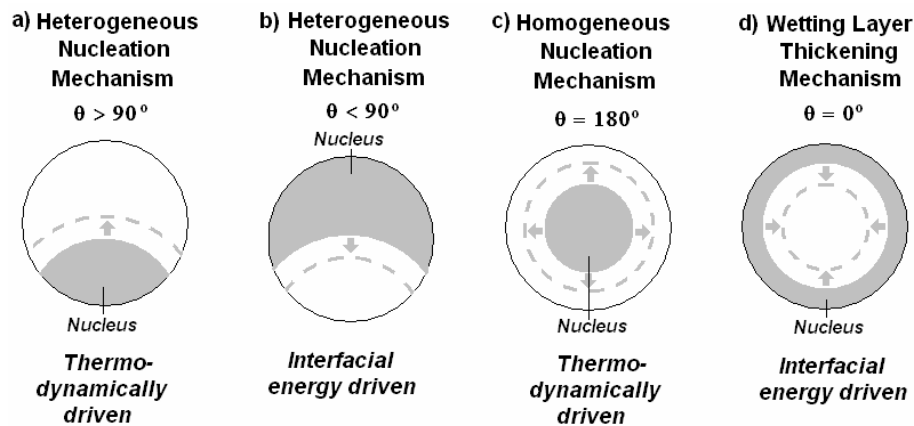

*FIG 7. Schematic diagram showing phase transformation mechanisms that can occur in confined volumes. All mechanisms can be modelled using equations (10) and (12).*

If the heterogeneous crystallization of a confined phase occurs with a contact angle of $\theta_c = \mathrm{acos}[(\gamma_1 - \gamma_2) / \gamma]$ then melting of this same system can also be expected to occur by a heterogeneous nucleation mechanism for which the contact angle, $\theta_m \sim \mathrm{acos}[(\gamma_2 - \gamma_1)/\gamma]$, i.e. $\sim 180° - \theta_c$. $\theta_m$ will be exactly $180° - \theta_c$ if the nucleus structure is equivalent to that of the new phase produced within the droplet, and so this will be more likely as $R$ approaches $r^*$. Hence we can predict the melting and crystallization temperatures upon and within substrates readily, if we know $\gamma$ and either $\theta_c$ or $\theta_m$. Figures 8a and b shows the predicted variation in ice crystallization and melting temperatures with substrate radius for the case of $\theta_c = 80°$, $\theta_m = 100°$, and $\theta_c = 100°$, $\theta_m = 80°$, respectively.

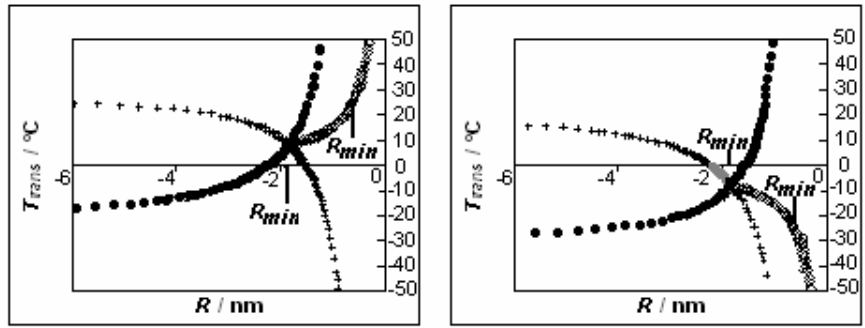

FIG. 8. *The predicted ice crystallization (filled circles) and melting temperatures (crosses) as a function of droplet radius, R, for heterogeneous crystallization within droplets. a) $\theta_c = 80°$, $\theta_m = 100°$, and b) $\theta_c = 100°$, $\theta_m = 80°$. Note that for $R < R_{min}$, the ice melting and crystallization temperatures are both given by the curve labelled $R_{min}$.*

### D. Phase transitions within small confined volumes

The graphs in Figures 8a and b show that the crystallization and melting curves for a particular system cross at $T_{trans} > T_0$ if $\theta_c < 90°$ and at $T_{trans} < T_0$ if $\theta_c > 90°$. We denote the droplet radius at which the crossing occurs as $R_{min}$. For all droplet radii smaller than $R_{min}$, we then have the unphysical situation that the melting curve is below the crystallization one. This occurs because a fundamental thermodynamic criterion is now being violated. In particular, the crystallization and melting curves for $R < R_{min}$ correspond to systems where the complete phase transformation of the droplet results in an increase in the free energy of the system, which of course cannot occur globally across the sample. This arises because, although the critical nucleus size can be attained as $\Delta G^*_{het}$ is surmountable, there is then insufficient material within the droplet for the free energy change to decrease to zero on further nucleus growth. In fact the limiting criterion that the free energy change must not be greater than zero has long been associated [7] with the minimum possible melting temperature of a small object. Here we are just extending this idea by also identifying the same criterion with the maximum possible freezing temperature of a confined object [3].

The effect of imposing the limiting condition that the free energy change must not be greater than zero on the total phase transformation of the droplet is easily obtained through the following. We set $R = R_{min}$ when $\Delta G_{het} = 0$ on complete phase transformation, and retain the convention that this radius takes a negative value for nucleation on a concave surface. Then,

$$\frac{4\pi R_{min}^3 \Delta\mu}{3v} + 4\pi R_{min}^2 (\gamma_2 - \gamma_1) = 0$$

i.e.
$$R_{min} = \frac{3v(\gamma_1 - \gamma_2)}{\Delta\mu} = \frac{3v\gamma}{\Delta\mu}\cos\theta = \frac{3r^*}{2}\cos\theta \qquad (17)$$

Equation (17) shows that $R_{min}$ only takes the required negative values for phase transformations within spherical substrates when $\theta > 90°$ and $r^*$ is positive, or when $\theta < 90°$ and $r^*$ is negative. Hence the $R_{min}$ condition occurs at $T_{trans} < T_0$ for $\theta > 90°$ and at $T_{trans} > T_0$ for $\theta < 90°$, as expected. For substrate sizes below $R_{min}$, $T_c$ and $T_m$ are no longer given by equations (13) and (14), but by equation (17) and equation (4), i.e.

$$T_{trans} = T_0\left[1 - \frac{3\gamma v \cos\theta}{R\Delta_{trans}H}\right]. \qquad (18)$$

So in Figure 8, for $R < R_{min}$, the ice melting and crystallization temperatures are both given by the curve labelled $R_{min}$. Hence, we would expect the hysteresis normally observed upon heating and cooling the same system to disappear for phase transformations confined to within volumes with $R \leq R_{min}$. As required, equation (18) reduces to the lower bound of the melting temperature when $\theta = 0$ and a pre-existing liquid or crystalline surface layer is present. A similar dependence of $\Delta T \propto 1/R$ would also be expected to apply for the limiting criterion that $\Delta G_{het} = 0$ on complete phase transformation within any closed concave-shaped system, but again use of a geometric factor would be required for the correct dependence, e.g. for a cylindrical vessel, $\Delta T = (2\gamma v T_0 \cos\theta)/R\Delta_{fus}H$ [5].

Please note that equation (18) assumes that $\gamma$, $\theta$ and $\Delta_{fus}H$ are invariant with $R$, but for very small droplet sizes, this will not necessarily be the case. However, the variation of $\gamma$, $\theta$ and $\Delta_{fus}H$ from their bulk values can be ascertained by measuring the extent to which the gradient of a $\ln R$ versus $\ln\Delta T$ plot deviates from -1, since for $R \leq R_{min}$,

$\frac{d\ln\Delta T}{d\ln R} = -1 + \frac{d}{d\ln R}\ln\left\{\frac{\gamma_1 - \gamma_2}{\Delta_{fus}H}\right\}$. Also for homogeneous nucleation, and other cases where the contact angle value is known, we can evaluate how the ratio $\gamma/\Delta_{fus}H$ differs from its bulk value [14]. This provides an independent measure of how highly-curved, nanoscale systems perturb the bulk ratio values of interfacial tensions to enthalpies of crystallization. Consequently, by using equation (12) for droplets with $R \geq R_{min}$, and equation (18) for droplets with $R \leq R_{min}$, the entire $T_c$ and $T_m$ versus $R$ dependence can be modelled. These equations can be used to model the melting and crystallization temperature variation with confinement volume for any system, within the limitations of our model, i.e. the assumption of incompressible phases, no volume change on phase transformation and isotropic nature of the critical nucleus, confining volume and interfacial tensions. For a constant $\theta_c > 90°$, we expect a drop off in crystallization temperatures for sufficiently small confinement volumes, which is observed in many systems [25,26]. The drop-off rate will be mediated, though, by any change in the contact angle and deviation of the ratio $\gamma/\Delta_{fus}H$ from its bulk value. In contrast, for a constant $\theta_c < 90°$, we expect the far rarer case of an increase in crystallization temperatures for sufficiently small confinement volumes, as has been observed for $CCl_4$ freezing in microporous activated carbon fibers [27]. We have recently applied [14] our model to the case of water freezing in microemulsions and have found additionally that it provides a simple and direct method of obtaining the critical nucleus size for the first time, provided the contact angle and droplet size are determinable. This is possible since when $R \leq R_{min}$, equation (17) $r^* = 2R/(3\cos\theta)$ holds, and so the critical nucleus size can be obtained directly without reliance on the Gibbs-Thomson equation and the inappropriate usage of bulk interfacial tension values in small confined volumes.

### III. CONCLUSIONS

Homogeneous phase transformations, and phase transformations driven by thickening of pre-existing surface layers are limiting cases of the more general heterogeneous nucleation theory. Hence we propose that values for crystallization and melting temperatures can be obtained by a simple extension to classical nucleation theory. This model predicts that crystallization and melting temperatures are given by roots of the equation, $\left(\Delta G^*_{hom} f(\theta_p)/kT_{trans}\right) = \ln A$, provided that the new phase grows to a size greater than $R_{min}$, where $R_{min} = 1.5 r^* \cos\theta$. For sizes below $R_{min}$, the requirement that the free energy change on complete phase transformation is not greater than zero means that melting and crystallization temperatures are then given by $T_{trans} = T_0 \left[1 - (3v\gamma\cos\theta / \Delta_{trans} H R)\right]$, and we would expect the hysteresis normally observed upon crystallizing and melting the same system to disappear. This simple model provides more accurate values for melting and crystallization temperatures in confined volumes than those obtained by using the Gibbs-Thomson equation or the unmodified classical nucleation theory.

## ACKNOWLEDGEMENTS

We thank N. Clarke for helpful discussions. This work was supported by a grant from the Engineering and Physical Sciences Research Council.

**APPENDIX**:

Proof that

$$f(\theta_p) = \frac{1 + 2x^3 - 3x^2 \cos\theta + y(1 + x\cos\theta - 2x^2)}{2} = \frac{1 - \cos^3(\theta + \phi) + 4x^3 f(\phi) - 3x^2 \cos\theta(1 - \cos\phi)}{2}$$

Since: $\cos\phi = \dfrac{x - \cos\theta}{y}$, $\cos^3(\theta + \phi) = \dfrac{(x\cos\theta - 1)^3}{y^3}$

and $4x^3 f(\phi) = x^3 \left(1 - \dfrac{x - \cos\theta}{y}\right)^2 \left(2 + \dfrac{x - \cos\theta}{y}\right) = 2x^3 + \dfrac{2x^3(\cos\theta - x)}{y} + \dfrac{x^3 \sin^2\theta(\cos\theta - x)}{y^3}$,

then,

$$0.5\left[1 - \cos^3(\theta + \phi) + 4x^3 f(\phi) - 3x^2 \cos\theta(1 - \cos\phi)\right]$$

$$= 0.5\left[1 - 3x^2 \cos\theta + 2x^3 + \dfrac{2x^3(\cos\theta - x) + 3x^2 \cos\theta(x - \cos\theta)}{y} - \dfrac{(x\cos\theta - 1)^3 - x^3 \sin^2\theta(\cos\theta - x)}{y^3}\right]$$

$$= 0.5\left[1 - 3x^2 \cos\theta + 2x^3 + \dfrac{-2x^4 + 5x^3 \cos\theta - 3x^2 \cos^2\theta}{y} - \dfrac{(x\cos\theta - 1)^3 - x^3 \sin^2\theta(\cos\theta - x)}{y^3}\right].$$

But:

$$(x\cos\theta - 1)^3 = (x^2 \cos^2\theta - 2x\cos\theta + 1)(x\cos\theta - 1) = (y^2 - x^2 \sin^2\theta)(x\cos\theta - 1)$$

$$= x\cos\theta y^2 - y^2 - x^3 \sin^2\theta \cos\theta + x^2 \sin^2\theta$$

so

$$(x\cos\theta - 1)^3 - x^3 \sin^2\theta(\cos\theta - x) = x\cos\theta y^2 - y^2 - x^3 \sin^2\theta \cos\theta + x^2 \sin^2\theta - x^3 \sin^2\theta \cos\theta + x^3 \sin^2\theta$$

$$= x\cos\theta y^2 - y^2 + x^4 \sin^2\theta - 2x^3 \sin^2\theta \cos\theta + x^2 \sin^2\theta = x\cos\theta y^2 - y^2 + x^2 \sin^2\theta y^2.$$

Hence

$$0.5\left[1 - \cos^3(\theta + \phi) + 4x^3 f(\phi) - 3x^2 \cos\theta(1 - \cos\phi)\right]$$

$$= 0.5\left[1 - 3x^2 \cos\theta + 2x^3 + \dfrac{-2x^4 + 5x^3 \cos\theta - 3x^2 \cos^2\theta - x\cos\theta + 1 - x^2 \sin^2\theta}{y}\right]$$

$$= 0.5\left[1 - 3x^2 \cos\theta + 2x^3 + \dfrac{-2x^4 + 5x^3 \cos\theta - 2x^2 \cos^2\theta - x\cos\theta + 1 - x^2}{y}\right]$$

$$= 0.5\left[1 - 3x^2 \cos\theta + 2x^3 + \frac{y^2(1 - 2x^2 + x\cos\theta)}{y}\right] = 0.5\left[1 - 3x^2 \cos\theta + 2x^3 + y(1 + x\cos\theta - 2x^2)\right]$$

<div align="right">QED</div>

## FOOTNOTES AND REFERENCES